\documentclass[10pt,letterpaper,twocolumn]{article} 

\usepackage{ol2}
\usepackage[draft]{hyperref}
\usepackage{amsmath}

\begin{document}

\twocolumn[ 

\title{Spatially single-mode source of bright squeezed vacuum}


\author{A.~M.~P\'erez$^{1,*}$, T.~Sh.~Iskhakov$^1$, P.~Sharapova$^{2,3}$, S.~Lemieux$^4$, O.~V.~Tikhonova$^{2,3}$, M.~V.~Chekhova$^{1,2,5}$, G.~Leuchs$^{1,5}$}

\address{
$^1$Max-Planck Institute for the Science of Light, \\  Guenther-Scharowsky-Str. 1 / Bau 24, Erlangen 91058, Germany \\
$^2$Physics Department, Lomonosov Moscow State University, Moscow 119991, Russia
\\
$^3$Skobeltsyn Institute of Nuclear Physics, Lomonosov Moscow State University, Moscow 119234, Russia
\\
$^4$Department of Physics, University of Ottawa, Ottawa, Ontario, Canada
\\
$^5$Universit\"at Erlangen-N\"urnberg, Staudtstrasse 7/B2, Erlangen 91058, Germany
\\
$^*$Corresponding author: angela.perez@mpl.mpg.de
}

\begin{abstract} Bright squeezed vacuum, a macroscopic nonclassical state of light, can be obtained at the output of a strongly pumped non-seeded traveling-wave optical parametric amplifier (OPA). By constructing the OPA of two consecutive crystals separated by a large distance we make the squeezed vacuum spatially single-mode without a significant decrease in the brightness or squeezing.

\end{abstract}


 ] 

\noindent Bright squeezed vacuum (BSV) is a state of light emerging at the output of an unseeded high-gain optical parametric amplifier. Although squeezed vacuum can be generated in cavities below threshold or in nonlinear waveguides, the states with the highest brightness (photon number per mode) have been so far achieved only in bulk crystals pumped by short pulses with high peak power~\cite{Bondani,two-color,single-mode}. This way of generation, however, leads to an essentially multimode structure. In this work, we show that by generating BSV in two spatially separated crystals one can achieve spatially single-mode structure of the radiation without considerable reduction of the brightness or loss of the nonclassical features.

A traveling-wave OPA made of two consecutive crystals with the optic axes tilted oppositely has been used before in order to increase twice the interaction length without the increase in the transverse walk-off~(see, for instance, Ref.~\cite{two-crystal} or recent Ref.~\cite{two-color}). It considerably reduces the effect of anisotropy on the generation of both two-photon light and BSV~\cite{anis}. Now, we will consider the two crystals separated by a larger distance.
\begin{figure} [h]
\centerline{\includegraphics[width=5cm]{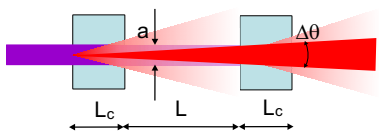}}
\caption{The idea of the experiment.}
\label{sketch}
\end{figure}
Let BSV be generated in two crystals of length $L_c$, separated by a distance of $L$, from a strong pump with the beam waist $a$ (Fig.~\ref{sketch}). Each crystal, taken separately, would emit parametric down-conversion (PDC) radiation into a rather broad solid angle (shown as transparent cones in the figure). However, if the parametric gain is high, then the PDC generation is strongly nonlinear, and the width of the angular spectrum $\Delta\theta$ is given by the part of the PDC radiation emitted by the first crystal and amplified in the second one. At $L\gg L_c$, the angle $\Delta\theta$ will be given by the pump beam waist and the distance between the crystals, $\Delta\theta=a/L$.

As a result, by increasing the distance between the crystals one can reduce the total angular width $\Delta\theta$ of the PDC spectrum. At the same time, the width of the photon-number correlations within the angular spectrum $\delta\theta$ is determined, roughly, by the pump angular divergence~\cite{Fedorov2007} and, as long as $L$ is less than the Rayleigh length for the pump, does not depend on the distance between the crystals. As soon as $\Delta\theta$ becomes equal to $\delta\theta$, the radiation becomes spatially single-mode. In our experiment, this is tested by measuring the second-order normalized intensity correlation function (CF) at zero time delay $g^{(2)}(0)$ after narrowband frequency selection at a non-degenerate wavelength. Because a single mode of PDC has thermal statistics, with $g^{(2)}(0)=2$~\cite{single-mode,Tapster,Bondani2004}, the measured value of the CF $g^{(2)}(0)_{meas}$ depends on the total number of collected modes $m$ according to the relation  $g^{(2)}(0)_{meas}=1+1/m$~\cite{Ivanova}. Here, $m$ includes both the number $m_l$ of longitudinal (frequency) modes and the number $m_t$ of transverse (angular) modes, $m=m_tm_l$. With narrowband spectral filtering, we can achieve $m_l=1.25$~\cite{single-mode}, and the value of $g^{(2)}(0)_{meas}$ can be used to infer the number of transverse modes.

\begin{figure} [h]
\centerline{\includegraphics[width=6cm]{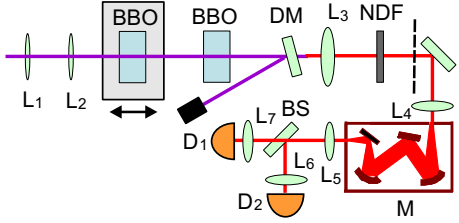}}
\caption{The experimental setup. BSV from two BBO crystals, after the elimination of the pump with a dichroic mirror DM, is collected into the input slit of a monochromator filtering out $1.25$ longitudinal modes. Further, $g^{(2)}(0)$ is measured to find out the number of transverse modes. Neutral-density filters (NDF) prevent the saturation of the detectors or a CCD (placed at the position of the dashed line).}
\label{setup}
\end{figure}

The experimental setup is shown in Fig.~\ref{setup}.
Third-harmonic radiation of a YAG:Nd laser, with the wavelength $355$ nm, pulse duration $18$ ps, repetition rate $1$ kHz and mean power $60$ mW, is softly focused by a telescope (lenses $L_{1,2}$) into a waist of $200$ $\mu m$. PDC is obtained from two $3$ mm BBO crystals cut for type-I collinear frequency-degenerate phase matching, the second crystal fixed after the beam waist and the first one placed on a movable platform. The platform can be displaced without any angular tilting in such a way that the distance between the crystals is changed from $7$ mm to $170$ mm. After the crystals, the pump radiation is eliminated by a dichroic mirror (DM). Due to the high value of the parametric gain, most part of the PDC radiation is produced in the second crystal. To collimate it, a lens ($L_3$) with a focal distance of $200$ mm is placed at $200$ mm from the second crystal. In its focal plane (shown by a dashed line), in its turn, a CCD camera can be placed (not shown) to capture the angular distribution of the PDC intensity. The camera is preceded by an interference filter (not shown) cutting a bandwidth of $10$ nm out of the PDC spectrum around the degenerate wavelength $709$ nm. In the absence of the camera, all radiation is collected by lens $L_4$ into the input slit of a monochromator (M) selecting a bandwidth of $0.1$ nm, which is sufficient for obtaining $m_l=1.25$~\cite{single-mode}. After the monochromator, the radiation is collimated with another lens ($L_5$) and fed, by means of a beamsplitter and two lenses $L_{6,7}$, to two charge-integrating detectors based on p-i-n diodes, and both integral intensity and normalized second-order intensity CF are measured. To prevent saturation of the camera and detectors, neutral density filters (NDF) are used in some measurements.

\begin{figure} [h]
\centerline{\includegraphics[width=5cm]{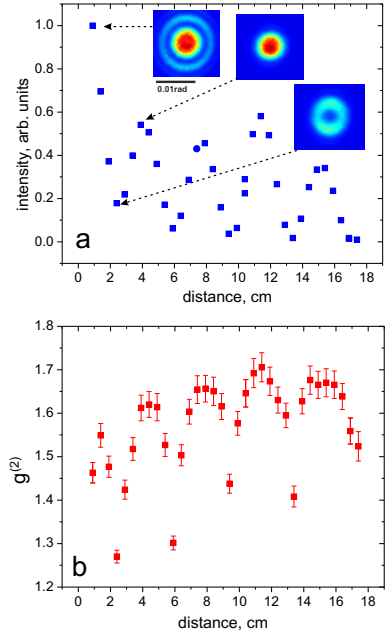}}
\caption{PDC intensity (a) and the second-order CF (b) versus the distance between the crystals. The insets show two-dimensional angular spectra of PDC recorded for different distances between the crystals.}
\label{expdata}
\end{figure}

In the first run, we measured the intensity of BSV and the normalized CF (Fig.~\ref{expdata}). As the distance $L$ between the crystals increased, both the integral intensity and the second-order normalized CF showed oscillations with the period $35$ mm. These oscillations are caused by the interference between the contributions to PDC from both crystals~\cite{big,DNK}.
 One can see that the peaks of the $g^{(2)}(L)$ dependence became higher with the growth of $L$. The highest value achieved was $1.71\pm0.03$. The maxima of $g^{(2)}$ corresponded to the maxima of intensity, and the shots made by the camera show that the corresponding angular spectra had nearly Gaussian shapes (insets to Fig.~\ref{expdata}a). At these points of the dependence, constructive interference between the PDC contributions from both crystals occurred. In contrast, the angular spectra at the minima had ring shapes, with the intensity at the center suppressed. These are points of destructive interference. They correspond to a larger number of modes due to their delocalized shapes. The total number of transverse modes varied from $2.96$ to $1.13$.

For a quantitative theoretical description of the effect, we used the formalism of Bloch-Messiah reduction~\cite{Bloch-Messiah}. The Hamiltonian was diagonalized by passing from plane waves to a new set of collective angular modes like the "broadband modes" introduced in~\cite{Christ} for the frequency domain. At high gain, the eigenvalues (Schmidt coefficients) are redistributed similar to Ref.~\cite{Wasilewski}: the strongest ones survive, the weaker ones are suppressed.

The total Hamiltonian describing high-gain PDC can be written in the form
\begin{eqnarray}
H=\int d\overrightarrow{k_s}\hbar\omega_s(a^\dagger_{k_s}a_{k_s}+\frac{1}{2})+\nonumber\\
+\int d\overrightarrow{k_i} \hbar\omega_i(a^\dagger_{k_i}a_{k_i}+\frac{1}{2})+\nonumber\\
+( i\hbar\Gamma\int d\overrightarrow{k_s} d\overrightarrow{k_i} F(\overrightarrow{k_s},\overrightarrow{k_i})a^\dagger_{k_s}a^\dagger_{k_i}+h.c.),
\label{Ham}
\end{eqnarray}
where the indices  \textit{s} and \textit{i} label the signal and idler photons respectively, $a^\dagger_{k_s},a^\dagger_{k_i}$ are the photon creation operators in modes $\overrightarrow{k_s},\overrightarrow{k_i}$, and $F(\overrightarrow{k_s},\overrightarrow{k_i})$ stands for the amplitude of the PDC process. Since we would like to describe only the spatial features of BSV, we assume the frequency phase-matching condition to be fulfilled: $\omega_s+\omega_i=\omega_p$. In this case, the interaction term in the Hamiltonian (\ref{Ham}) can be written as
\begin{equation}
H_{nl}\propto i\hbar\Gamma\int d\theta_s d\theta_i F(\theta_s,\theta_i)a^\dagger_{\theta_s}a^\dagger_{\theta_i}+h.c.,
\end{equation}
where $\theta_s,\theta_i$ are the angles of emission for signal and idler photons, respectively, and $a^\dagger_{\theta_s},a^\dagger_{\theta_i}$ are the photon creation operators in the corresponding plane-wave modes. The amplitude $F(\theta_s,\theta_i)$ for the two-crystal configuration can be calculated using the approach of Ref.~\cite{DNK}, which yields
\begin{eqnarray}
F(\theta_s,\theta_i)\propto \exp\lbrace -\frac{\sigma^2k^2_p(\theta_s+\theta_i)^2}{8}\rbrace \mathrm{sinc}\lbrace\frac{L_c k_p(\theta_s-\theta_i)^2}{16}\rbrace\nonumber\\
\times\cos\lbrace\frac{(L_c+\frac{L}{n_p})k_p(\theta_s-\theta_i)^2}{16}- \frac{\delta k L}{2 n_p}\rbrace.
\label{TPA}
\end{eqnarray}
Here,  $ 2\sqrt{ln 2}\sigma$ is the full width at half maximum (FWHM) of the pump spatial intensity distribution, $k_p$ and $n_p$ are the wave vector and the refractive index of the pump in the crystal, and the additional mismatch $\delta k=\frac{\omega_p n_p}{c}\delta n_{air}$  depends on the difference between the refractive indices of the pump and the signal and idler waves in the air space gap: $\delta n_{air}\equiv n_p-\frac{n_s+n_i}{2}$. According to (\ref{TPA}), $F(\theta_s,\theta_i)$ as a function of $L$ oscillates with the period of approximately $35$ mm, which is caused by the difference of refractive indices $\delta n_{air}=1.016\cdot10^{-5}$.

To obtain the photon numbers in the signal or idler channels and the correlations between them as a function of $L$, we develop an analytical approach based on the collective angular Schmidt modes. According to the Schmidt decomposition, the amplitude $F(\theta_s,\theta_i)$ can be represented as
\begin{equation}
F(\theta_s,\theta_i)=\sum_n\sqrt{\lambda_n}u_n(\theta_s)v_n(\theta_i),
\label{Schmidt}
\end{equation}
where $\lambda_n$ are the eigenvalues and $u_n(\theta_s)$, $v_n(\theta_i)$, the eigenfunctions of the reduced density matrix of the bipartite system. Using decomposition (\ref{Schmidt}), we introduce new creation/annihilation operators in the Schmidt modes as
\begin{equation}
A_n^\dagger=\int d\theta_s u_n(\theta_s)a^\dagger_{\theta_s},\,
B_n^\dagger=\int d\theta_i v_n(\theta_i)a^\dagger_{\theta_i}.
\label{op}
\end{equation}
Spatial collective modes were also introduced in \cite{Brambilla2004,Brambilla2008} but their evolution could be only obtained by numerical solution of integro-differential equations. In the Heisenberg picture, the evolution of operators (\ref{op}) is given by Hamilton's equations, which in our case can be solved analytically:
\begin{eqnarray}
\frac{d A_n}{dt}=\Gamma\sqrt{\lambda_n}B_n^\dagger-i\omega_s A_n ,\nonumber\\
\frac{d B_n}{dt}=\Gamma\sqrt{\lambda_n}A_n^\dagger-i\omega_i B_n.
\label{Heis}
\end{eqnarray}
Using the connection between the Schmidt and plane-wave operators we obtain the explicit expressions for the latter as functions of time and calculate, for different distances \textit{L} between crystals, both the integral photon numbers in the signal and idler beams, $\langle N_{s,i} \rangle$ and the second-order autocorrelation function $g^{(2)}$,
\begin{equation}
g^{(2)}=\frac{\langle N^{2}_{s}\rangle}{\langle N_{s}\rangle^2},
\end{equation}
where the averaging is performed over the initial vacuum state of the system. It should be mentioned that in order to obtain the correct dependence of $g^{(2)}$ on the distance \textit{L} we perform the analysis both for non-anisotropic and anisotropic directions. In the latter case, partial compensation of anisotropy due to the opposite tilts of the optic axes of the two crystals is taken into account by using the expression for $F(\theta_s,\theta_i)$ introduced in Ref.~\cite{anis}. The final value of $g^{(2)}$ is calculated as $g^{(2)}=1+\frac{1}{K}$ where the Schmidt number \textit{K} represents the effective total number of independent Schmidt modes and the number of frequency Schmidt modes is $1.25$.

\begin{figure} [h]
\centerline{\includegraphics[width=5cm]{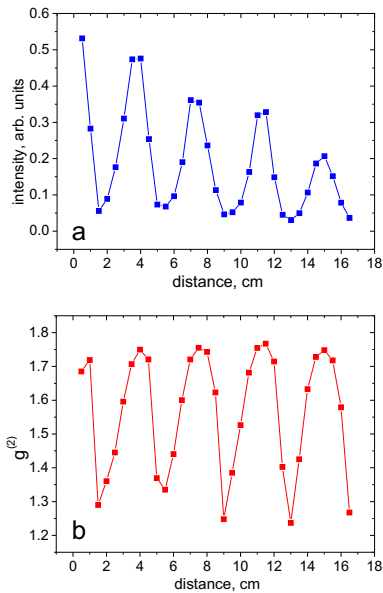}}
\caption{Calculated intensity in the signal channel (a) and the second-order CF (b) versus the distance between the crystals. The lines are guides to the eye.}
\label{theory}
\end{figure}

The results of the calculation are presented in Fig.~\ref{theory}.
The periodic modulation of both the intensity and the CF, as well as the envelope growth of the CF, are reproduced. Note that after reaching its peak value, the CF starts to decrease, and the same dependence is observed in the experiment. This is explained by the fact that although at each point of constructive interference (peak of CF) there is a strong pronounced nearly-Gaussian maximum, there is also a background caused by the non-interfering part of PDC radiation. At large $L$, the maxima become reduced as the part of the radiation amplified in the second crystal becomes narrower than the first eigenmode. Then the background gets a higher relative weight, and the number of effective modes increases. This occurs after the fourth peaks in Figs.~\ref{expdata},\ref{theory}. For the same reason, the highest value of CF is never reached for the current gain values. Also, note that the theory predicts a monotonic reduction of the envelope for the intensity dependence on $L$ while the experiment shows an increase of intensity at $L=11$ cm. This is because in experiment, this position corresponded to the waist of the beam while the theory assumed a constant beam diameter everywhere.

\begin{figure} [h]
\centerline{\includegraphics[width=5cm]{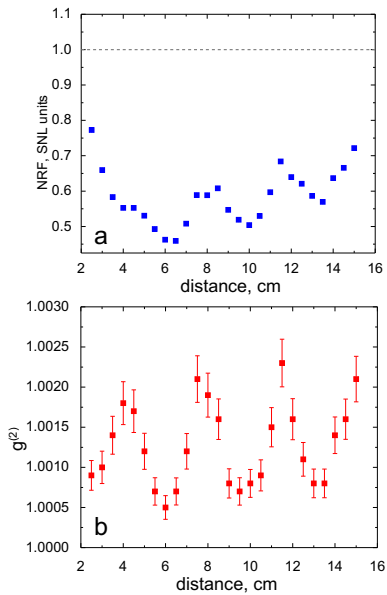}}
\caption{NRF (a) and CF (b) measured without the frequency filtering versus the distance between the crystals. Dotted straight line: the shot-noise level.}
\label{squeez}
\end{figure}

It is important that even with the crystal separated, we were able to observe considerable amount of squeezing. For this, we re-aligned the crystal for frequency non-degenerate phasematching with the wavelengths $635$ nm and $805$ nm, replaced the beamsplitter by a dichroic mirror and removed the monochromator with the lenses. The resulting setup was similar to the one for two-color squeezing measurement~\cite{two-color}, with the only exception that all angular spectrum was collected. Squeezing was characterized by the noise reduction factor (NRF) defined as the variance of the photon-number difference normalized to the mean value of the photon-number sum. The results of NRF measurement are presented in Fig~\ref{squeez}. In the same plot, the value of the normalized intensity auto-correlation function is shown. Although it is very close to unity due to the presence of many longitudinal modes, its dependence on the distance between the crystals is qualitatively the same as in Fig.~\ref{expdata}.
One can see that the maxima of the CF, i.e., the cases of fewer modes in the angular spectrum, correspond to the peaks of NRF and therefore to reduced squeezing. This is because for fewer angular modes, the setup becomes more critical to the alignment and, in particular, to the presence of unmatched modes in the two channels~\cite{two-color}. At the same time, a certain mismatch is always present as long as the signal and idler beams differ in wavelength, because the angle $\Delta\theta$ is selected by purely geometrical relations while in two-color squeezing measurements, the angular widths selected in the two channels should scale as the corresponding wavelengths~\cite{two-color}. Nevertheless, even in the case of single-mode generation ($L=11.5$ cm), considerable squeezing (NRF=0.65) was observed.

In conclusion, by constructing a traveling-wave parametric amplifier of two spatially separated nonlinear crystals, we have obtained bright squeezed vacuum with nearly single spatial mode ($m_t=1.13$) and the angular spectrum close to Gaussian. Single-mode generation was achieved at the expense of a certain reduction in both intensity and squeezing; however, photon-number correlation considerably below shot-noise level was still achieved. Further improvement of the noise reduction in such a configuration can be obtained by reducing the wavelength difference.

The research leading to these results received funding from the EU FP7 under grant No.
308803 (project BRISQ2) and ERA-Net.RUS (project Nanoquint). We also acknowledge partial financial support of the Russian Foundation for Basic Research, grants 12-02-00965, 14-02-00389-a, and 14-02-31084-mol-a.


\begin{thebibliography}{99}

\bibitem{Bondani}M.~Bondani, A.~Allevi, G.~Zambra, M.~G.~A.~Paris, and A.~Andreoni,
Phys. Rev. A  \textbf{76}, 013833(1)-013833(4) (2007).

\bibitem{two-color}I.~N.~Agafonov, M.~V.~Chekhova, and G.~Leuchs, PRA \textbf{82}, 011801 (2010).

\bibitem{single-mode} T.~Sh.~Iskhakov, A.~M.~Perez, K.~Yu.~Spasibko, M.~V.~Chekhova, and G.~Leuchs, Optics Letters \textbf{37}, 1919 (2012).

\bibitem{two-crystal}P.~Grangier, R.~E.~Slusher, B.~Yurke, and A.~LaPorta, PRL \textbf{59}, 2153 (1987).

\bibitem{anis} A.~M.~Perez, F.~Just, A.~Cavanna, M.~V.~Chekhova, and G.~Leuchs, Laser Physics Letters \textbf{10}, 125201 (2012).

\bibitem{Fedorov2007} M.~V.~Fedorov, M.~A.~Efremov, P.~A.~Volkov, E.~V.~Moreva, S.~S.~Straupe, and S.~P.~Kulik, Phys. Rev. Lett. \textbf{99}, 063901-1--063901-4 (2007).

\bibitem{Tapster} P.~R.~Tapster and J.~G.~Rarity, J. Mod. Opt. \textbf{45}, 595--604 (1998).

\bibitem{Bondani2004} F.~Paleari, A.~Andreoni, G.~Zambra, and M.~Bondani, Optics Express, \textbf{12}, Issue 13, 2816--2824 (2004).

\bibitem{Ivanova}O.~A.~Ivanova, T.~Sh.~Iskhakov, A.~N.~Penin, and M.~V.~Chekhova, Quantum Electronics \textbf{36}, 951--956 (2006).

\bibitem{big} A.~V.~Burlakov, M.~V.~Chekhova, D.~N.~Klyshko, S.~P.~Kulik, A.~N.~Penin, Y.~H.~Shih, and D.~V.~ Strekalov, PRA \textbf{56}, 3214--3225 (1997).

\bibitem{DNK} D.~N.~Klyshko, JETP , \textbf{105}, 1574 (1994).

\bibitem{Bloch-Messiah} A.~M.~Perez, P.~Sharapova, O.~V.~Tikhonova, M.~V.~Chekhova, and G.~ Leuchs, to be submitted (2014).

\bibitem{Christ} Andreas Christ, Kaisa Laiho, Andreas Eckstein, Katiuscia N Cassemiro, and Christine Silberhorn, New Journ. of Phys. \textbf{13}, 033027 (2011).

\bibitem{Wasilewski} W.~Wasilewski, A.~I.~Lvovsky, K.~Banaszek, and C.~Radzewicz, PRA \textbf{73}, 063819 (2006).

\bibitem{Brambilla2004} E.~Brambilla, A.~Gatti, M.~Bache, and L.~A.~Lugiato, Phys Rev A \textbf{69}, 023802 (2004).

\bibitem{Brambilla2008} E.~Brambilla, L.~Caspani, O.~Jedrkiewicz, L.~A.~Lugiato, and A.~Gatti, Phys Rev A \textbf{77}, 053807 (2008).

\end{thebibliography}

\newpage
\begin{thebibliography}{99}

\bibitem{Bondani}M.~Bondani, A.~Allevi, G.~Zambra, M.~G.~A.~Paris, and A.~Andreoni,
\textquotedblleft Sub-shot-noise photon-number correlation in a mesoscopic twin beam of light\textquotedblright,
~Phys. Rev. A  \textbf{76}, 013833(1)-013833(4) (2007).

\bibitem{two-color}I.~N.~Agafonov, M.~V.~Chekhova, and G.~Leuchs, \textquotedblleft Two-color squeezed vacuum\textquotedblright,~PRA \textbf{82}, 011801 (2010).

\bibitem{single-mode} T.~Sh.~Iskhakov, A.~M.~Perez, K.~Yu.~Spasibko, M.~V.~Chekhova, and G.~Leuchs, \textquotedblleft Superbunched bright squeezed vacuum state\textquotedblright,~Optics Letters \textbf{37}, 1919 (2012).

\bibitem{two-crystal}P.~Grangier, R.~E.~Slusher, B.~Yurke, and A.~LaPorta, \textquotedblleft Squeezed-Light--Enhanced Polarization Interferometer\textquotedblright,~PRL \textbf{59}, 2153 (1987).

\bibitem{anis} A.~M.~Perez, F.~Just, A.~Cavanna, M.~V.~Chekhova, and G.~Leuchs, \textquotedblleft Compensation of anisotropy effects in a nonlinear crystal for squeezed vacuum generation\textquotedblright,~Laser Physics Letters \textbf{10}, 125201 (2012).

\bibitem{Fedorov2007} M.~V.~Fedorov, M.~A.~Efremov, P.~A.~Volkov, E.~V.~Moreva, S.~S.~Straupe, and S.~P.~Kulik, ``Anisotropically and High Entanglement of Biphoton States Generated in Spontaneous Parametric Down-Conversion", Phys. Rev. Lett. \textbf{99}, 063901-1--063901-4 (2007).

\bibitem{Tapster} P.~R.~Tapster and J.~G.~Rarity, ``Photon statistics of pulsed parametric light", J. Mod. Opt. \textbf{45}, 595--604 (1998).

\bibitem{Bondani2004} F.~Paleari, A.~Andreoni, G.~Zambra, and M.~Bondani, ``Thermal photon statistics in spontaneous parametric downconversion", Optics Express, \textbf{12}, Issue 13, 2816--2824 (2004).

\bibitem{Ivanova}O.~A.~Ivanova, T.~Sh.~Iskhakov, A.~N.~Penin, and M.~V.~Chekhova, \textquotedblleft Multiphoton correlations in parametric down-conversion
and their measurement in the pulsed regime\textquotedblright,~Quantum Electronics \textbf{36}, 951--956 (2006).


\bibitem{big} A.~V.~Burlakov, M.~V.~Chekhova, D.~N.~Klyshko, S.~P.~Kulik, A.~N.~Penin, Y.~H.~Shih, and D.~V.~ Strekalov, ``Interference effects in spontaneous two-photon parametric scattering
from two macroscopic regions", PRA \textbf{56}, 3214--3225 (1997).

\bibitem{DNK} D.~N.~Klyshko, ``Interference effects ...", JETP , \textbf{105}, 1574 (1994).

\bibitem{Bloch-Messiah} A.~M.~Perez, P.~Sharapova, O.~V.~Tikhonova, M.~V.~Chekhova, and G.~ Leuchs, ``Schmidt modes in the angular spectrum of bright squeezed vacuum", to be submitted (2014).

\bibitem{Christ} Andreas Christ, Kaisa Laiho, Andreas Eckstein, Katiuscia N Cassemiro, and Christine Silberhorn, \textquotedblleft Probing multimode squeezing with
correlation functions\textquotedblright,~New Journ. of Phys. \textbf{13}, 033027 (2011).

\bibitem{Wasilewski} W.~Wasilewski, A.~I.~Lvovsky, K.~Banaszek, and C.~Radzewicz, \textquotedblleft Pulsed squeezed light: Simultaneous squeezing of multiple modes\textquotedblright, PRA \textbf{73}, 063819 (2006).

    \bibitem{Brambilla2004} E.~Brambilla, A.~Gatti, M.~Bache, and L.~A.~Lugiato, \textquotedblleft Simultaneous near-field and far-field spatial quantum correlations in the high-gain regime
of parametric down-conversion\textquotedblright, Phys Rev A \textbf{69}, 023802 (2004).

\bibitem{Brambilla2008} E.~Brambilla, L.~Caspani, O.~Jedrkiewicz, L.~A.~Lugiato, and A.~Gatti, \textquotedblleft High-sensitivity imaging with multi-mode twin beams\textquotedblright, Phys Rev A \textbf{77}, 053807 (2008).









\end{thebibliography}
\end{document}